\documentclass[aps, preprint,showpacs,preprintnumbers,amsmath,amssymb,nofootinbib,longbibliography]{revtex4-1}
\usepackage[colorlinks=true, pdfstartview=FitV, linkcolor=red, citecolor=blue, urlcolor=black, pdftitle={},pdfauthor={Yuki Minami, Yoshimasa Hidaka},pdfsubject={}, pdfkeywords={}]{hyperref}

\usepackage{amsmath,amssymb,bm}
\usepackage[dvips]{graphicx}
\usepackage{yfonts}
\usepackage{enumerate}
\usepackage{bm}
\usepackage{color}

\newcommand{\phir}{\phi_R}
\newcommand{\pir}{\pi_R}
\newcommand{\phia}{\phi_A}
\newcommand{\pia}{\pi_A}

\newcommand{\hphi}{{\phi}}
\newcommand{\hpi}{{\pi}}
\newcommand{\hphir}{{\phi}_R}
\newcommand{\hpir}{{\pi}_R}
\newcommand{\hphia}{{\phi}_A}
\newcommand{\hpia}{{\pi}_A}
\newcommand{\D}{\mathcal{D}}
\newcommand{\hQ}{{Q}}
\newcommand{\hH}{{H}}
\newcommand{\hbmq}{{\bm{q}}}
\newcommand{\hbmp}{{\bm{p}}}

\newcommand{\hvarphi}{\phi}
\newcommand{\Qr}{\hQ_R^{\alpha}}
\newcommand{\Qa}{\hQ_A^{\alpha}}
\newcommand{\Qmr}{\hQ_R^\alpha}
\newcommand{\Qma}{\hQ_A^\alpha}
\newcommand{\T}{[T^\alpha]^a_{~b}}
\newcommand{\vphi}{\phi}
\newcommand{\ri}{\mathrm{i}}

\begin{document}

\preprint{RIKEN-QHP-198, RIKEN-STAMP-15}
\author{Yuki Minami}
\affiliation {
Department of Physics, Kyoto University, Kyoto 606-8502, Japan}
\author{Yoshimasa Hidaka}
\affiliation{
Theoretical Research Division, Nishina Center, RIKEN, Wako, Saitama 351-0198, Japan}
\affiliation{
iTHEMS Program, RIKEN, Wako, Saitama 351-0198, Japan}


\title{Spontaneous symmetry breaking and Nambu-Goldstone modes in dissipative systems}
\begin{abstract}
We discuss  spontaneous breaking of internal symmetry and its Nambu-Goldstone (NG) modes in dissipative systems.
We find that there exist two types of NG modes in dissipative systems corresponding to type-A and type-B NG modes in Hamiltonian systems.
To demonstrate the symmetry breaking, we consider a $O(N)$ scalar model obeying a Fokker-Planck equation.
We show that the type-A NG modes in the dissipative system are diffusive modes, while they are propagating modes in Hamiltonian systems. 
We point out that this difference is caused by the existence of two types of Noether charges, $\hQ_R^\alpha$ and $\hQ_A^\alpha$:
$\hQ_R^\alpha$ are symmetry generators of Hamiltonian systems, which are not conserved in dissipative systems.
$\hQ_A^\alpha$ are symmetry generators of dissipative systems described by the Fokker-Planck equation, which are conserved. 
We find that the NG modes are propagating modes if $\hQ_R^\alpha$ are conserved, while those are diffusive modes if they are not conserved.
We also consider a $SU(2)\times U(1)$ scalar model with a chemical potential to discuss the type-B NG modes.
We show that the type-B NG modes have a different dispersion relation from those in the Hamiltonian systems.
\end{abstract}
\maketitle

\section{Introduction}
\label{sec:intro}

Spontaneous symmetry breaking (SSB) is a universal phenomenon and widely observed 
 at various scales in nature, e.g., our vacuum where the electroweak and chiral symmetries are spontaneously broken,
superconductors, ferromagnets, and solid crystals~\cite{nambu1961dynamical, goldstone1961field, goldstone1962broken, higgs1964broken}.
Those  well known examples are in Hamiltonian systems. 

As is the case with the Hamiltonian systems, it is known that the SSB occurs even in dissipative systems such as reaction diffusion~\cite{pattern} and active matter~\cite{PhysRevLett.75.4326, PhysRevE.58.4828} systems, and synchronization transition of coupled oscillators~\cite{kuramoto2012chemical,RevModPhys.77.137}. 
The reaction diffusion system has translational symmetry. The symmetry is spontaneously broken by a pattern structure in space~\cite{pattern} such as the Turing pattern, which is considered as the most basic pattern formation in biology~\cite{turing1952chemical}.
In the active hydrodynamics, which describes a collective motion of biological organisms such as flocks of birds, the rotational symmetry is spontaneously broken.
Toner and Tu showed that the Nambu-Goldstone (NG) modes of the active hydrodynamics in $d$ dimensions are given as $d-2$ diffusive shear modes and a pair of propagating sound modes \cite{PhysRevLett.75.4326, PhysRevE.58.4828}.   
Recently, the propagating sound mode seems to be experimentally observed in the situation where a flock collectively turns~\cite{attanasi2014information}.
Attanasi et al. also phenomenologically discussed the type of the NG mode, propagating or diffusive modes, based on the SSB of the rotational symmetry and conservation law~\cite{attanasi2014information}.  
In the synchronization transition that describes the behavior of chemical and biological oscillators~\cite{kuramoto2012chemical}, U(1) phase symmetry is spontaneously broken by synchronization of the coupled oscillators.
Furthermore, a diffusion mode appears as the NG mode associated with the SSB of the U(1) symmetry. The NG mode has the different property from  the U(1) symmetry broken of Hamilton systems, where the mode is a propagating mode.

In Hamiltonian systems, NG modes 
are classified into two types: type-A and type-B~\cite{PhysRevLett.108.251602, PhysRevLett.110.091601, PhysRevX.4.031057,PhysRevD.91.056006}. 
When a global symmetry $\mathcal{G}$ is broken into its subgroup $\mathcal{H}$,
the numbers of type-A ($N_\text{A}$) and type-B ($N_\text{B}$) NG modes are expressed as
\begin{align}
N_\text{A} = N_\text{BS}-\mathrm{rank}\langle[\ri Q^\alpha,Q^\beta]\rangle,\qquad
N_\text{B} = \frac{1}{2}\mathrm{rank}\langle[\ri Q^\alpha,Q^\beta]\rangle, \label{eq:NGrelation}
\end{align}
where $N_\text{BS}=\mathrm{dim}(\mathcal{G}/\mathcal{H})$ is the number of broken symmetries, and  $Q^\alpha$ are the Noether charges (generators) of $\mathcal{G}$.
The total number of the NG modes is $N_\text{A}+N_\text{B}=N_\text{BS} - \mathrm{rank}\langle[\ri Q^\alpha,Q^\beta]\rangle/2$~\cite{PhysRevLett.108.251602, PhysRevLett.110.091601, PhysRevX.4.031057,PhysRevD.91.056006,Takahashi:2014vua}.
For the spontaneous breaking of an internal symmetry, both type-A and type-B NG modes are propagating with the linear and quadratic dispersions, respectively.

Compared to those in Hamiltonian systems, the relation among the SSB, the NG modes and their dispersion relations in dissipative systems are not fully understood.
We cannot naively apply the above argument to the dissipative systems because it is based on conservation laws.
In the dissipative systems, the symmetry does not mean that the generators of the symmetry are the conserved quantities. For example, a Brownian particle obeying a Fokker-Planck equation has a rotational symmetry; however, the angular momentum is not conserved due to the dissipation and noise. 
Nevertheless, we know that the SSB occurs and there appear gapless modes, in the absence of  the conservation law. 
In the above example of the active hydrodynamics, the rotational symmetry is spontaneous broken by the expectation value of the nonvanishing velocity field,
where the angular momentum is not conserved. 

To discuss the symmetry breaking  in the dissipative systems, it is useful to consider a path integral formulation of a Langevin or Fokker Planck equation.
The formalism is  called the Martin-Siggia-Rose (MSR) formalism~\cite{PhysRevA.8.423, Pre}, which is successful in the analysis of dynamic critical phenomena~\cite{siggia1976renormalization, RevModPhys.49.435, PhysRevB.18.353,Täuber20127}.
Even in the dissipative system, we can construct the Noether charge $\Qa$, which is the conserved quantity, as the generator of the symmetry in the Lagrangian of MSR formalism.  This Noether charge is a different charge from that in the Hamiltonian system. We refer the Noether charge in the Hamiltonian system to $\Qr$, which is not conserved in the dissipative systems.

In this paper, we show that the SSB in dissipative systems can be discussed by $\Qa$ instead of $\Qr$.
To this end, we will consider a model with $O(N)$ scalar fields $\hphir^a(t,\bm{x})$.
When $O(N)$ symmetry is spontaneous broken into $O(N-1)$, nonvanishing order parameters $\langle[\ri \Qa,\hphir^a(t,\bm{x})]\rangle$ appear.
In this case, the NG modes corresponding to the type-A NG modes become diffusive modes. 
The existence of diffusive modes is the characteristic feature in the dissipative systems because the NG modes associated with spontaneous breaking of an internal symmetry in the Hamiltonian systems are propagating modes.
We also consider a $SU(2)\times U(1)$ scalar model with a chemical potential.
We find that a type-B NG mode appears in the dissipative system. 
The mode has the quadratic dispersion as in the Hamiltonian system, but the damping rate has different order in momentum. 
In both examples, the absence of the conservation of $\Qr$  is essential to determine the patterns of dispersion relations.
Furthermore, as a system with a nonequilibrium stationary state, we consider a model that exhibits synchronization and corresponds to a SU(2) $\times$ U(1) version of the complex Gintzburg-Landau equation~\cite{RevModPhys.65.851}.
Then, we show that there exists type-A and -B NG modes even in the nonequilibrium stationary state.
Finally, we nonperturbatively establish these results by using a Ward-Takahashi identity  for $\Qa$ and $\Qr$,
which cover not only thermal equilibrium states but also nonequilibrium steady states.

This paper is organized as follows.
In Sec.~\ref{sec:brownian}, to see symmetry in dissipative systems, we consider a rotational symmetry of Brownian motion as a simple example,
in which we discuss the mathematical similarity between  Fokker-Planck and  Schr\"odinger equations.
We also give the MSR formalism for readers who are unfamiliar with.
 In Sec.~\ref{sec:ssb}, we consider $O(N)$ and $SU(2)\times U(1)$ models to discuss the spontaneous breaking of internal symmetries and their NG modes in dissipative systems. 
In Sec.~\ref{sec:ward}, we establish the results in Sec.~\ref{sec:ssb} in the model independent way using the Ward-Takahashi identity.
Section~\ref{sec:summary} is devoted to the summary and discussion.
In Appendix~\ref{sec:keldysh}, 
we  discuss an action in the real-time formalism, which has two types of Noether charges, $\Qr$ and $\Qa$.  The action reduces to that in a dissipative system by a coupling to an environment, which violates the $\Qr$ symmetry.

\section{Rotational symmetry in dissipative systems and Martin-Siggia-Rose formalism}
\label{sec:brownian}
We briefly review that symmetry of a Langevin equation can be discussed as in operator and  path integral formalisms in quantum mechanics~\cite{fong2013noether, graham1985integrability, ovchinnikov2013transfer, misawa1994new, gaeta1999lie}.
To this end, we consider a rotational symmetry of a Brownian motion.
The Langevin equation for the Brownian particle $\bm{x}(t)$ is given by
\begin{align}
\frac{d}{dt}{\bm{x}}(t)&=\bm{u}(t),  \label{eq:brown0}\\
\frac{d}{dt} \bm{u}(t)&=-\gamma \bm{u}(t) + \bm{\xi}(t),  \label{eq:brown}
\end{align}
where $\bm{u}(t)$ is the velocity,
 $\gamma$ the friction coefficient and 
$\bm{\xi}(t)$ the random noise that satisfies the fluctuation-dissipation relation,
\begin{align}
\langle \xi_i (t) \xi_j(t')\rangle = 2\gamma T \delta_{ij}\delta(t-t'). \label{eq:FDR}
\end{align}
Here, $\langle ... \rangle$ represents the average over the noise, and $T$ is the temperature of a heat bath.
In this section, we set the mass of the Brownian particle to unity without loss of generality.

If $\bm{\xi}$ and $\gamma$ vanish, the system reduces to the Hamiltonian system, and the angular momentum 
$\bm{L}_R=\bm{x}\times \bm{u}$
is conserved. This results from the rotational symmetry of the equation of motion, $\bm{x}\to \bm{x}'=R\,\bm{x}$ and $\bm{u}\to \bm{u}'=R\,\bm{u}$  with a rotation matrix $R$.
The angular momentum plays a role of the generator of the rotational symmetry, $\{x_i,L_{Rj}\}_\text{PB}=\epsilon_{ijk}x_k$, where $\epsilon_{ijk}$ is the Levi-Civita tensor, and $\{... , ...\}_{\text{PB}} $ represents the Poisson bracket, $\{x_i,u_j\}_\text{PB}=\delta_{ij}$.
In contrast, when $\bm{\xi}$ and $\gamma$ exist, the angular momentum is no longer conserved, $d{\bm{L}}_R/dt= -\gamma \bm{x}\times\bm{u} + \bm{x}\times\bm{\xi}\neq0$ due to  the friction and the noise. However, this does not mean that the absence of the rotational symmetry. In fact, 
Eqs.~\eqref{eq:brown0}-\eqref{eq:FDR} are still rotationally symmetric under $\bm{x}\to \bm{x}'=R\,\bm{x}$, $\bm{u}\to \bm{u}'=R\,\bm{u}$, and $\bm{\xi}\to \bm{\xi}'=R \,\bm{\xi}$.
As we will see in the following,  this rotational symmetry implies that the existence of another conserved quantity.

For this purpose, it is useful to introduce the probability distribution of the velocity,
\begin{align}
P(\bm{v},t) \equiv \langle \delta^{(3)}(\bm{u}(t)-\bm{v})\rangle.
\end{align}
The time evolution of $P(\bm{v},t)$ obeys  the Fokker-Planck equation,
\begin{align}
\partial_t P (\bm{v},t)=  \biggl( 
\gamma T \frac{\partial^2}{\partial \bm{v}^2}+\gamma\frac{\partial}{\partial \bm{v}}  \bm{v}
   \biggr) P(\bm{v},t). \label{eq:fpeq2}
\end{align}
Here, a point is that we can regard  the Fokker-Planck equation as a Schr\"odinger equation~\cite{justin1989quantum}.
If we rewrite  
\begin{align}
\bm{v} \rightarrow \hbmq, \quad \frac{\partial}{\partial \bm{v}} \rightarrow \ri\hbmp,
\end{align}
which naturally satisfy the commutation relation $[{q}_i, {p}_j ] = \ri\delta_{i j}$,
the Fokker-Planck equation is expressed as 
\begin{align}
\partial_t | P (t) \rangle= -\hH_\text{FP} |P(t) \rangle \label{eq:fpeq}
\end{align}
with the Hamiltonian,
\begin{align}
\hH_\text{FP}= \gamma T \hbmp^2-\ri\gamma \hbmp \cdot \hbmq.  \label{eq:Hamiltonian}
\end{align}
$| P (t) \rangle$ is the state vector, and $\hH_\text{FP}$ is the Fokker-Planck Hamiltonian.
The Fokker-Planck equation~\eqref{eq:fpeq2} is identified as the coordinate representation 
of Eq.~\eqref{eq:fpeq}. 
The important difference from quantum mechanics is that the Hamiltonian \eqref{eq:Hamiltonian} is not hermitian, and thus, the left and right eigenstates are generally different.

Obviously, the Hamiltonian $\hH_\text{FP}$ is invariant under $\hbmq\to R\hbmq$ and $\hbmp\to R\hbmp$, so that we find the ``angular momentum'' as the Noether charge of the rotational symmetry,
\begin{align}
{\bm{L}}_A= \hbmq \times \hbmp,
\end{align}
which commutes with the Fokker-Planck Hamiltonian $\hH_\text{FP}$.
We emphasize  that ${\bm{L}}_A$ is not the actual angular momentum because ${\bm{L}}_A={-} \ri \bm{v} \times (\partial / \partial \bm{v})\neq\bm{L}_R (=\bm{x}\times \bm{q})$. 
The role of  ${\bm{L}}_A$ is the generator of rotational symmetry;
for example, the commutation relation between ${L}_{Ai}$ and ${q}_j$ $(p_{i})$ gives 
\begin{align} 
[{L}_{Ai}, {q}_j]=\ri \epsilon_{ijk}{q}_k, \qquad  [{L}_{Ai}, {p}_j]=\ri \epsilon_{ijk}{p}_k.
\end{align}
These are compared with the commutation relation with $L_{Ri}$,
\begin{equation}
\begin{split}
[L_{Ri}, q_{j}] = 0, \qquad  [L_{Ri}, p_{j}] = \ri \epsilon_{ijk}{x}_k.
\end{split}
\end{equation}
In contrast to $\bm{L}_A$, $\bm{L}_{R}$ does not commute with the Fokker-Planck Hamiltonian, $[ \hH_\text{FP},\bm{L}_{R} ]= -\gamma \bm{x}\times(2\ri T\bm{p}+\bm{q})\neq0$.

We can also consider the rotational symmetry of states. 
Let us consider the following state $| \psi (\bm{q}_0) \rangle$ as an example:
\begin{align}
\langle \bm{q} | \psi (\bm{q}_0) \rangle = N \exp \biggl[-\frac{1}{2}(\bm{q}-\bm{q_0})^2\biggr],
\end{align}
where $ \langle \bm{q} |$ is the left eigenstate of $\hbmq$ 
and $N$ is the normalization constant.
By operating ${\bm{L}}_A$ to $| \psi (\bm{q}_0) \rangle$, we obtain
\begin{align}
\langle \bm{q} | {\bm{L}}_A |\psi (\bm{q}_0) \rangle ={-}\ri N(\bm{q} \times \bm{q_0} )\exp \biggl[-\frac{1}{2}(\bm{q}-\bm{q_0})^2\biggr].
\end{align}
That is, we have
\begin{align}
 {\bm{L}}_A |\psi (\bm{q}_0) \rangle &= 0 \text{ for the symmetric state}, \notag\\
 &\neq 0 \text{ for the non-symmetric state}, \notag
\end{align} 
as in quantum mechanics.
Thus, we can map the Langevin equation to that in the ``quantum mechanics'' and define the ``Hamiltonian'' and the ``Noether charge.'' 
If a stationary solution of the Fokker-Planck equation~\eqref{eq:fpeq} is a non-symmetric state, the symmetry is spontaneous broken.
By using these, we can discuss the symmetry of dissipative systems and its spontaneous breaking as in quantum field theories.
We note that the expectation value of $\bm{L}_{A}$ always vanishes due to conservation of the probability~\cite{justin1989quantum, 0034-4885-79-9-096001}.
It is rather useful to introduce an order parameter defined as the expectation value of commutator of charge and an operator, $-\ri\langle [\bm{L}_{A}, O]\rangle$,
which is often employed in quantum field theories.

It is also useful to introduce the path-integral representation of the generating functional $Z[\bm{J}]$ called 
the Martin-Siggia-Rose (MSR) formalism~\cite{justin1989quantum, PhysRevA.8.423},
\begin{align}
Z[\bm{J}]&\equiv \langle e^{\ri \int dt \bm{J}(t) \cdot \bm{q}(t)}\rangle \notag\\
&=\int \D \bm{q} \D \bm{p} \mathcal{J}e^{{\ri}S[\bm{q}, \bm{p}]+\ri \int dt \bm{J}(t)\cdot \bm{q}(t)}, \label{eq:browngeneratingfunction}
\end{align}
where $\bm{J}$ represents the source and $\mathcal{J} \equiv \det (\partial_t+\gamma)$ is the jacobian. Here, we drop the jacobian since it is independent of $\bm{q}$ and $\bm{p}$.
The action $S[\bm{q},\bm{p}]$ is expressed as
\begin{align}
\ri S[\bm{q},\bm{p}] &=\int dt [  \ri\bm{p} \cdot \partial_t{\bm{q}}-H_\text{FP} ] \notag\\
&=\int dt \biggl[ \ri\bm{p} \cdot (\partial_t\bm{q}+\gamma \bm{q}) -\gamma T \bm{p}^2\biggr]  \notag\\
 &=\frac{1}{2}\int dt 
\begin{pmatrix}
\bm{q} & \bm{p}
\end{pmatrix}
\begin{pmatrix}
0 & \ri D_{A}^{-1} \\
\ri D_{R}^{-1} & -2\gamma T  
\end{pmatrix}
\begin{pmatrix}
\bm{q} \\ \bm{p}
\end{pmatrix}.
\end{align} 
In the last line, we symmetrized  $\bm{p} \cdot \partial_t{\bm{q}}$ term as 
$(\bm{p} \cdot \partial_t{\bm{q}} - \bm{q} \cdot \partial_t{\bm{p}})/2$, and introduced the inverse of the retarded and advanced Green functions 
$D_{R,A}^{-1}$ as 
\begin{align}
D_{R,A}^{-1} = \pm \partial_t + \gamma.
\end{align}
The MSR action is useful to discuss the SSB in the dissipative systems.
Namely, we can use standard techniques for the SSB even in the dissipative systems 
as we shall see the next section.

\section{Spontaneous symmetry breaking in dissipative systems}
\label{sec:ssb}
In this section, we consider SSB in dissipative systems.
We first provide a toy model of scalar fields, $\phi^a(t,\bm{x})$, with $O(N)$ symmetry.
We discuss the symmetry and its Noether charge based on the MSR formalism.
We shall find that the Noether charge that we call $\Qa$ is obtained from the symmetry of the action in the MSR formalism. $\Qa$ corresponds to ${\bm{L}}_A$  of the Brownian particle in the previous section.
In addition, if we drop the dissipation from our model, 
another Noether charge $\Qr$ arises, which corresponds to the actual angular momentum of the Brownian particle, $\bm{L}_R$. 

Next, we discuss the spontaneous breaking of $O(N)$ symmetry in the language of quantum field theories. The corresponding nonvanishing order parameters are $\langle[\ri \Qa,\phi^a(t,\bm{x})]\rangle$. In our model, we will have all $\langle[\ri \Qa,Q_A^\beta]\rangle=\langle[\ri \Qa,Q_R^\beta]\rangle=0$, so that the NG modes belong to type-A NG modes.
Furthermore, we show that the NG modes of $O(N)$ model become diffusive modes.
This behavior is quite different from those in the Hamiltonian system, in which the NG modes become propagating modes,
such as the phonon in superfluids. That is, the diffusive NG mode is the characteristic of dissipative systems.
In Sec.~\ref{sec:SU(2)xU(1)model}, we also consider a $SU(2)\times U(1)$ model with a chemical potential, which is a simple model for type-B NG mode.
In this model, we will have a nonvanishing order parameter, $\langle[\ri Q_A^\alpha,Q_R^\beta]\rangle$, and thus, there appear a type-B NG mode.
We shall find the type-B mode has a different dispersion relation from that in the Hamiltonian system.
In Sec.~\ref{sec:SU(2)xU(1)modelDriven}, we discuss a toy model that exhibit a synchronization transition as symmetry breaking in a nonequilibrium stationary state. 
The stationary states of the toy models in Secs.~\ref{sec:SSBO(N)model} and~\ref{sec:SU(2)xU(1)model} are the thermal equilibrium states although the dissipation violates conservation laws.
We will show that the behaviors of the dispersion relation for type-A and -B NG modes do not change even in the nonequilibrium stationary state. 
 
\subsection{$O(N)$ scalar model and its symmetry}
\label{sec:model}
Let us consider the following Langevin equation as a toy model:
\begin{align}
\partial_t \phi^a_R(t, \bm{x} ) -\{ \phi^a_R(t, \bm{x} ),  F\}_{\text{PB}} &= 0, \label{eq:langevin1-1}\\
\partial_t \pi^a_R(t, \bm{x}) -\{ \pi^a_R(t, \bm{x} ),  F  \}_{\text{PB}} + \gamma \frac{\delta F}{\delta \pi^a_R} &= \xi^a(t, \bm{x}), 
\label{eq:langevin1-2}
\end{align} 
where $\phir^a$ and $\pir^a$ are the scalar fields that belong to the fundamental representation of $O(N)$ symmetry, the subscript $a$ runs $1, 2, ..., N$, $\gamma$ is the dissipation constant, 
and $\{... , ...\}_{\text{PB}} $ represents the Poisson bracket:
\begin{align}
\{ X,  Y \}_{\text{PB}} \equiv\int d^3x\biggl[ \biggl(\frac{\partial X}{\partial \phir^a(\bm{x})}\biggr)\biggl( \frac{\partial Y}{\partial \pir^a(\bm{x})}\biggr) -  \biggl( \frac{\partial Y}{\partial \phir^a(\bm{x})} \biggr)\biggl(\frac{\partial X}{\partial \pir^a(\bm{x})} \biggr) \biggr].
\end{align}
$\xi^a(t, \bm{x})$ is the random noise satisfying
\begin{align}
\langle \xi^a(t,\bm{x}) \xi^b(t', \bm{x}')\rangle&=A \delta^{ab}\delta(t-t')\delta^{(3)}(\bm{x}-\bm{x}'),
\end{align}
where $A$ represents the strength of the noise. If we assume that the fluctuation-dissipation relation, $A=2\gamma T$.
$F$ is the free energy, which has the following form: 
\begin{align}
F[\phi_R,\pi_R] = \int d^3{x} \biggl[\frac{1}{2}(\pi_R^a)^2
+\frac{1}{2}(\bm{\nabla}\phi^a_R)^2+\frac{1}{2}m^2(\phi_R^a)^2
+\frac{u^2}{4}((\phi_R^a)^2)^2\biggr].
\label{eq:freeenergy}
\end{align}
If we drop $\gamma$ and $\xi^a$, 
our model reduces to the usual Hamiltonian equation whose Hamiltonian is given by Eq.~\eqref{eq:freeenergy}.   
Hence, without the dissipation, our model turns out to be the classical version of the Goldstone model, which is the simplest model for the SSB and the NG modes
\cite{ goldstone1961field}.

The corresponding MSR action and Fokker-Planck Hamiltonian read
\begin{align}
\ri S &=\int {d^4x}  \biggl[ \ri\pia^a\biggl( \partial_t \phir^a - \pir^a\biggr) \notag\\
&\qquad-\ri\phi^a_A \biggl( \partial_t \pir^a +\gamma \pir^a 
+\biggl( -\bm{\nabla}^2+m^2+u^2(\phir^b)^2 \biggr)\phir^a \biggr) 
-\frac{A}{2}(\phia^a)^2
\biggr], \label{eq:action}
\end{align}
and
\begin{align}
\hH_\text{FP}=\int {d^3x} \biggl[\ri\hpi_A^a \hpi_R^a
+\ri\hphi^a_A \biggl( \gamma \hpi_R^a 
+\biggl( -\bm{\nabla}^2+m^2+u^2({\hphi}_R^b)^2 \biggr){\hphi}_R^a \biggr) +\frac{A}{2}({\hphi}_A^a)^2\biggr],
\end{align}
where ${\hpi}_A^a$ and ${\hphi}_A^a$ are the canonical momentum, which satisfy the commutation relations,
\begin{align}
[{\hphi}_R^a(t,\bm{x}), {\hpi}_A^b(t,\bm{x}')]&=[{\hphi}_A^a(t,\bm{x}),{\hpi}_R^b(t,\bm{x}')]=\ri \delta^{ab}\delta^{(3)}(\bm{x}-\bm{x}'), 
\end{align}
and the others are zero. 
In Eq.~\eqref{eq:action}, we have dropped the terms coming from the jacobian, which is independent of the fields in our model, and therefore it does not affect results as in the Brownian motion Eq.~(\ref{eq:browngeneratingfunction}). 

The stationary solution of this Fokker-Planck equation is the Gibbs distribution,
\begin{align}
\langle \phi_R,\pi_R|P\rangle=P(\phi_R,\pi_R) = \frac{1}{Z} e^{-F[\phi_R,\pi_R]/T},
\end{align}
where $T\equiv A/(2\gamma)$ is the temperature, and $Z$ is the normalization constant such that $\int d\phi_R d\pi_R P(\phi_R,\pi_R)=1$.
$|P\rangle$ is the right eigenstate of $\hH_\text{FP}$ with the zero eigenvalue.

The action~\eqref{eq:action} is invariant under under the following infinitesimal $O(N)$ transformations,
\begin{align}
\phir^a &\rightarrow \phir^a+\ri \epsilon_{\alpha} \T \phir^b, \quad 
\pir^a \rightarrow \pir^a+\ri \epsilon_{\alpha} \T \pir^b, \\
\phia^a &\rightarrow \phia^a+\ri \epsilon_{\alpha} \T \phia^b, \quad 
\pia^a \rightarrow \pia^a+\ri \epsilon_{\alpha} \T \pia^b,
\end{align}
where  $\epsilon_\alpha$ is an infinitesimal parameter, and $\T$ is the generator of $O(N)$ group, which satisfies 
the Lie algebra, $[T^\alpha, T^{\beta}] = \ri f^{\alpha\beta\gamma} T^\gamma$
with the structure constant $f^{\alpha\beta\gamma}$.
For example, the generator of $O(3)$ symmetry is given as
\begin{align}
T^1 = 
\begin{pmatrix}
0 & 0 & 0 \\
0 & 0 & -\ri \\
0 & \ri & 0 \\
\end{pmatrix}, \quad
T^2 = 
\begin{pmatrix}
0 & 0 & \ri \\
0 & 0 & 0 \\
-\ri & 0 & 0 \\
\end{pmatrix},\quad
T^3 = 
\begin{pmatrix}
0 & -\ri & 0 \\
\ri & 0 & 0 \\
0 & 0 & 0 \\
\end{pmatrix}.
\end{align}
From this symmetry of the action, we obtain the Noether charge,
\begin{align}
\hQ_A^\alpha=-\int d^3{x} \biggl[ {\hpi}_A^a \ri \T {\hphi}_R^b 
                 + {\hpi}_R^a \ri \T {\hphi}_A^b \biggr]. \label{eq:Qma}
\end{align}
$\hQ_A^\alpha$ corresponds to $\bm{{L}}_A$ of the Brownian motion in Sec.~\ref{sec:brownian}.
$\hQ_A^\alpha$ satisfies the Lie algebra, $[\hQ_A^\alpha,\hQ_A^\beta] = \ri f^{\alpha\beta\gamma} \hQ_A^\gamma
$.
If $\gamma=A=0$, i.e.,  the Hamiltonian system, 
the action~\eqref{eq:action} is invariant under another infinitesimal transformations:
\begin{align}
{\phia}^a \rightarrow {\phia}^a+\ri \epsilon_\alpha \T \phir^b, \quad
\pia^a \rightarrow \pia^a+\ri \epsilon_\alpha \T \pir^b,
\label{eq:rtransformation}
\end{align}
and $\phir^a \rightarrow \phir^a,  
\pir^a \rightarrow \pir^a$.
The Noether charge of this symmetry is given as 
\begin{align}
\hQ_R^\alpha=-\int d^3{x} 
\biggl[  {\hpi}_R^a \ri \T {\hphi}_R^b \biggr].
\label{eq:Qmr}
\end{align}
$\Qmr$ corresponds to  the actual angular momentum in Sec.~\ref{sec:brownian}.
In fact, their Poisson bracket satisfies the Lie algebra, $ \{Q_{R}^\alpha,Q_{R}^\beta\}_{\text{PB}}= f^{\alpha\beta\gamma}Q_R^\gamma$, although their commutation relation does not, $[\hQ^\alpha_R,\hQ^\beta_R]=0$\footnote{In quantum field theories, $[\hQ^\alpha_R,\hQ^\beta_R]$ may be nonzero. See Appendix~\ref{sec:keldysh}.}.
In the MSR formalism, the doubling of the fields, $\phi_R$ and $\phi_A$, occurs and it causes the doubling of the symmetry for the Hamiltonian system.
We will further discuss the doubling in the real-time (or the Keldysh) formalism,  which describes a real-time evoluation of quantum fields, in Appendix~\ref{sec:keldysh}. 
In the real-time formalism, a field doubles as time-forward and –backward ones, and a generator of a symmetry also doubles.
The two generators rotate the time-forward and –backward fields, respectively. Therefore, the doubling of the field leads to that of the generators, $\Qr$ and $\Qa$. These two symmetries are natural in quantum systems~\cite{0034-4885-79-9-096001}.

We also give the action by integrating  $\pir^a$ and $\pia^a$ out:
\begin{align}
\ri S = \int {d^4x}\biggl[
-\frac{1}{2}
\begin{pmatrix}
\phir^a & \phia^a
\end{pmatrix}
\begin{pmatrix}
0 & \ri D_{A}^{-1} \\
\ri D_{R}^{-1} & A  
\end{pmatrix}
\begin{pmatrix}
\phir^a \\ \phia^a
\end{pmatrix}
- \ri u^2 (\phir^b)^2 \phia^a\phir^a \biggr],
\label{eq:msraction2}
\end{align}
where $d^4x\equiv dt d^3{x} $ and 
\begin{align}
D_{R,A}^{-1} =\partial_t^2\pm\gamma\partial_t-\bm{\nabla}^2 +m^2. \label{eq:DRA}
\end{align}
In the next section, we discuss the SSB in this model and the dispersion relation of the NG modes.

\subsection{Spontaneous symmetry breaking in a $O(N)$ model} \label{sec:SSBO(N)model}
Let us discuss the SSB in the $O(N)$ model with the MSR action~\eqref{eq:msraction2}.
When the squared mass is negative, $-\mu^2\equiv m^2<0$, the symmeteric state $\langle{\phi^a_i}\rangle=0$ is disfavored because 
the propagator obtained from Eq.~\eqref{eq:DRA} contains an unstable mode, $\sim e^{t|\mu|}$.
To find the true stable state, we look for the stationary solution of the MSR action~\eqref{eq:msraction2}.
For this purpose, we consider the following potential:
\begin{align}
V_{\rm eff}=m^2 \ri \phia^a\phir^a+\frac{A}{2}(\phia^a)^2+ u^2 (\phir^b)^2 \ri \phia^a\phir^a,
\end{align}
which is obtained from the action with homogenous fields.
The stationary solutions are given by
\begin{align}
\frac{\delta V_{\rm eff}}{\delta \phir^a} &
= \biggl(-{\mu^2} +u^2 (\phir^b)^2 \biggr)\ri \phia^a+2 u^2 \ri \phia^b\phir^b \phir^a=0, \\
\frac{\delta V_{\rm eff}}{\delta \phia^a} &
=\ri \biggl(-{\mu^2} +u^2 (\phir^b)^2 \biggr)\phir^a+A \phia^a=0,
\end{align}
and we obtain the two nontrivial solutions:
\begin{align}
(\phir^a)^2 &=\frac{{\mu^2}}{u^2}, \quad \text{and} \quad \phia^a=0,  \label{eq:firstSolution}\\
(\phir^a)^2 &={\frac{{\mu^2}}{3u^2}}, \quad \text{and} \quad \phia^a={\ri \frac{2 {\mu^2}}{3A} \phir^a}, \label{eq:secondSolution}
\end{align}
in addition to $\phir^a=\phia^a=0$ corresponding to the unstable symmetric state.
The second solution~\eqref{eq:secondSolution} is also unstable, whereas the first solution~\eqref{eq:firstSolution} is stable.
To see the instability of the second solution,
let us suppose $\phir^a$ at $a=1$ has the nonzero expectation value:
\begin{align}
\langle \phir^a\rangle = \frac{\phi_0}{\sqrt{3}}\delta^{a}_{1},
\label{eq:classicalSolution}
\end{align}
where $\phi_0 = {\mu} /u$.
The squared mass for $\phir^b$ for $b=2, 3, ..., N$ of the second solution is negative: 
\begin{align}
[D_R^{-1}]_{11}(\omega=0, \bm{k}=\bm{0})&=\left. \frac{\delta^2 V_{\rm eff}}{\ri \delta \phia^1 \delta \phir^1} \right|_{\phi_0/\sqrt{3}} 
=0, \\
[D_R^{-1}]_{bc}(\omega=0, \bm{k}=\bm{0})&=\left. \frac{\delta^2 V_{\rm eff}}{\ri \delta \phia^b \delta \phir^c} \right|_{\phi_0/\sqrt{3}} 
=-\delta^{bc}\frac{2}{3}\mu^2.
\end{align}
In contrast, for the first solution, we set  
$\langle \phir^a\rangle = \phi_0 \delta^{a}_1$,
and obtain the positive squared mass
\begin{align}
\left. \frac{\delta^2 V_{\rm eff}}{\ri \delta \phia^1 \delta \phir^1} \right|_{\phi_0}=2{\mu^2},  \quad \text{and}  \quad \left.
\frac{\delta^2 V_{\rm eff}}{\ri \delta \phia^b \delta \phir^c} \right|_{\phi_0}=0 .
\end{align} 
Therefore, the stable solution is  the first solution~\eqref{eq:firstSolution} and the SSB occurs.

As in quantum field theories, the symmetry breaking is characterized by a nonvanishing expectation value of a Noether charge and a local operator.
In our case, it is 
\begin{align}
 -\langle [\ri \Qma, \hphir^a] \rangle =\ri [T^\alpha]^a_{~1}\phi_0.
\end{align}
There are $(N-1)$'s independent nonvanishing components. Thus, the $O(N)$ symmetry is spontaneously  broken into the $O(N-1)$ symmetry.
The corresponding $\hphir^b$ for $b=2, 3, ..., N$ are the Nambu-Goldstone fields.
For example, in the case of $O(3)$,  $\hQ_{A}^{2}$, $\hQ_{A}^{3}$ and $\hphir^3$, $\hphir^2$ are the broken Noether charges and the NG fields, respectively:
 \begin{align}
-\langle [\ri \hQ_{A}^2, \hphir^3] \rangle = \langle [\ri \hQ_{A}^3, \hphir^2] \rangle
=  \phi_0\neq 0. \label{eq:ssbo3}
\end{align}
Furthermore, $\Qmr$ are also spontaneously broken if we take the limit $\gamma\to0$ and $A\to0$:
\begin{align}
-\langle [\ri \Qmr, \hphia^a] \rangle= \ri [T^\alpha]^a_{~1}\phi_0.
\end{align}
Since $\langle [\ri \Qmr, Q^\beta_A] \rangle=\langle [\ri \Qma, Q^\beta_A] \rangle=0$, the NG modes belong to type-A modes.

Now, we discuss the dispersion relation of NG modes.
To this end, we consider fluctuations around the expectation value \eqref{eq:classicalSolution},
in which we parametrize the fields as
\begin{align}
\phir^a(x)=( \phi_0 + \sigma_R(x), \chi_R^b(x) ), {\qquad \phia^a(x)=( \sigma_A(x), \chi_A^b(x) ),}
\end{align}
where the subscript $b$ runs $2, 3, ... ,N$.
The MSR action in $\sigma$ and $\chi^b$ turns out to be
\begin{align}
\ri S&= \int {d^4x} \biggl[
-\frac{1}{2}\begin{pmatrix}
\sigma_R & {\sigma_A}
\end{pmatrix}
\begin{pmatrix}
0 & \ri D_{\sigma, A}^{-1} \\
\ri D_{\sigma, R}^{-1} & A  
\end{pmatrix}
\begin{pmatrix}
\sigma_R \\ {\sigma_A}
\end{pmatrix}\notag\\
&\qquad-\frac{1}{2}
\begin{pmatrix}
\chi_R^b &{\chi_A^b}
\end{pmatrix}
\begin{pmatrix}
0 & \ri D_{\chi, A}^{-1} \\
\ri D_{\chi, R}^{-1} & A  
\end{pmatrix}
\begin{pmatrix}
\chi_R^b \\ {\chi_A^b}
\end{pmatrix}
-V_{\rm int}
\biggr],
\end{align}
where the inverse propagators $D_{\sigma, \chi}^{-1}$ and the interaction term $V_\text{int}$ are given as
\begin{align}
D_{\sigma, R, A}^{-1} &= \partial_t^2 \pm \gamma \partial_t -\bm{\nabla}^2+2u^2\phi_0^2, \\
D_{\chi, R, A}^{-1} &= \partial_t^2 \pm \gamma \partial_t -\bm{\nabla}^2, \\ \label{eq:ngmode}
V_{\rm int}&= \ri u^2\biggl[
\sigma_A\biggl(3\phi_0\sigma_R^2+\phi_0(\chi_R^b)^2+\sigma_R(\chi_R^b)^2 {+\sigma_R^3}\biggr)
+\chi_A^b\chi_R^b\biggl(2\phi_0\sigma_R+(\chi_R^b)^2+\sigma_R^2\biggr)
\biggr].
\end{align}
We can see that $\chi^b$ does not have the mass term, i.e., $\chi^b$ is gapless.
The dispersion relation of the type-A NG modes is determined by 
\begin{align}
D_{\chi, R}^{-1} (\omega, \bm{k})=-\omega^2-\ri \gamma\omega+\bm{k}^2=0 \label{eq:dispersion}.
\end{align}
The solutions are given as
\begin{align}
\omega (\bm{k}) &= -\frac{\ri}{2}\gamma \pm \frac{\ri}{2}\sqrt{\gamma^2 - 4 {\bm{k}}^2} \notag\\
                      &\sim -\frac{\ri}{\gamma}\bm{k}^2 \text{  and  }  
                                      - \ri \gamma+\frac{\ri}{\gamma} \bm{k}^2,
\end{align}
where we have expanded $\omega(\bm{k})$ up to the second order in $\bm{k}$. 
Since $\omega (\bm{k})$ has no real part, these modes are purely damping modes at small $\bm{k}$.
One is the diffusive mode in which the damping vanishes at $\bm{k}=\bm{0}$.
The other has a finite damping even at $\bm{k}=\bm{0}$. 
In this model, the number of diffusive NG modes coincides with the number of broken symmetries $N-1$.
We note that NG modes in Hamiltonian systems become propagating modes such as spin waves in ferromagnets~\cite{PhysRevD.91.056006}. 
In fact, if we set $\gamma=0$ in Eq.~\eqref{eq:dispersion}, we obtain the propagating mode, $\omega (\bm{k})= \pm |\bm{k}|$.
As mentioned in the Introduction, the diffusive NG mode is the characteristic of the dissipative system.

\subsection{Spontaneous symmetry breaking in a $SU(2)\times U(1)$ model} \label{sec:SU(2)xU(1)model}
Here, we discuss the dispersion relation of the type-B NG modes in a dissipative system.
We consider a $SU(2)\times U(1)$ model with a chemical potential, which is known as a simple model for realizing type-B NG modes~\cite{Miransky:2001tw,Schafer:2001bq}.
Suppose that  the MSR action has the form,
\begin{align}
\ri S&= \int {d^4x} \Bigl(
\ri \varphi_A^\dag((-(\partial_0+\ri \mu)^2+\bm{\nabla}^2-\gamma\partial_0)\varphi_R -2\lambda(\varphi_R^\dag\varphi_R)\varphi_R)\notag\\
&\quad+
\ri \varphi_R^\dag((-(\partial_0+\ri \mu)^2+\bm{\nabla}^2+\gamma\partial_0)\varphi_A -2\lambda(\varphi_R^\dag\varphi_R))\varphi_A
-A\varphi_A^\dag\varphi_A
\Bigr),
\end{align}
where $\varphi_{i}=(\varphi^1_{i},\varphi^2_{i})$ is the two component complex scalar fields,  $\mu$  the chemical potential, $\lambda$ the coupling constant, and $\gamma$ the friction coefficient.
This action is invariant under $SU(2)\times U(1)$ transformation, 
$\varphi_i \to \varphi_i+ \ri \epsilon_\alpha T^\alpha \varphi_i$,
where $T^0$ is the $U(1)$ generator, and $T^a$ ($a=1,2,3$) are the $SU(2)$ generators satisfying the Lie algebra, $[T^a,T^b]= \ri \epsilon^{abc} T^c$.
We choose the normalization of the generators as $\mathrm{tr} \, T^\alpha T^\beta = \delta^{\alpha\beta}/2$.
The Noether charges are given as
\begin{align}
\hQ_A^\alpha= -\int d^3{x} \Bigl[ {\hpi}_A^\dag \ri T^\alpha {\varphi}_R
                 + {\hpi}_R^\dag \ri T^\alpha {\varphi}_A
                  -{\varphi}_R^\dag \ri T^\alpha {\pi}_A
                 - {\varphi}_A^\dag \ri T^\alpha {\pi}_R
                  \Bigr],
                  \label{eq:QAB}
\end{align} 
where $\pi_R=(\partial_0+\ri \mu)\varphi_R$ and $\pi_A=(\partial_0+\ri \mu-\gamma)\varphi_A$.
We can also define 
\begin{align}
Q_R^{\alpha}= -\int d^3x\Bigl[ \pi_R^\dag \ri T^\alpha\varphi_R
- \varphi_R^\dag \ri T^\alpha\pi_R
\Bigr],
\label{eq:QRB}
\end{align}
which are the Noether charges associated with the transformation, $\varphi_A\to\varphi_A+\ri \epsilon_\alpha T^\alpha\varphi_R$.
Let us find the stationary solution of $\delta S/\delta \varphi_i=0$. Using the same analysis in Sec.~\ref{sec:SSBO(N)model}, 
we find a stable stationary solution, $\varphi_R=(0,v)$ with $v=\mu/\sqrt{2\lambda}$. 
Since the following expectation values,
\begin{align}
 -\frac{1}{V}\langle[\ri Q_A^1,Q_R^2]\rangle=\frac{1}{V}\langle[\ri Q_A^2,Q_R^1]\rangle=\mu v^2 \label{eq:typeBCondition}
\end{align}
are nonvanishing, $Q_R^1$ and $Q_R^2$ correspond to the type-B NG fields in the nondissipative limit. 
Here, $V$ is the volume of the system. 

To analyze the dispersion relation, we parametrize the fields as $\varphi_R=(\chi^1_R+\ri \chi^2_R, v+\psi^1_R+\ri \psi^2_R)$, and $\varphi_A=(\chi^1_A+\ri \chi^2_A,\psi^1_A+\ri \psi^2_A)$. Then, we find the inverse propagator for $\psi$ and $\chi$ sectors,
\begin{align}
D_{\psi R}^{-1}(\omega,\bm{k})&=
\begin{pmatrix}
-\omega^2-\ri \gamma\omega+\bm{k}^2+4\lambda v^2 &  2\ri \mu\omega\\\
-2\ri \mu\omega\ & -\omega^2-\ri \gamma\omega+\bm{k}^2
\end{pmatrix}, \label{eq:DinvpsiR}\\
D_{\chi R}^{-1}(\omega,\bm{k})&=
\begin{pmatrix}
-\omega^2-\ri \gamma\omega+\bm{k}^2 & 2\ri \mu\omega\\
-2\ri \mu\omega & -\omega^2-\ri \gamma\omega+\bm{k}^2
\end{pmatrix}.
\end{align}
The dispersion relation is obtained from $\det D_{\psi R}^{-1}=0$ and $\det D_{\chi R}^{-1}=0$.
At small $\bm{k}$, we find the diffusive NG mode $\omega =-\ri |\bm{k}|^2/\gamma$ in the $\psi$ sector and the type-B mode,
\begin{align}
\omega =  \frac{|\bm{k}|^2}{4\mu^2+\gamma^2}(\pm2\mu- \ri \gamma),
\end{align}
in the $\chi$ sector.
In the limit of $\gamma\to0$, we obtain the dispersion relations $\omega=\pm|\bm{k}|/\sqrt{3}$ and  $\omega=\pm |\bm{k}|^2/(2\mu)$ in the $\psi$ and $\chi$ sectors, respectively.
Therefore, in this model, the type-B NG mode is the propagating mode with the quadratic dispersion,  while the type-A NG mode is the diffusive mode.

\subsection{Driven dissipative condensate in a $SU(2)\times U(1)$ model } \label{sec:SU(2)xU(1)modelDriven}
In the previous examples, the stationary state is thermal equilibrium, although the system is dissipative. Here, we would like to consider a system with a nonequilibrium stationary state.
For this purpose, we introduce a complex coupling constant $\lambda=\lambda_{\mathrm{r}}+\ri\lambda_{\mathrm{i}}$ and a complex mass term $m_{\mathrm{r}}^{2}+\ri m^2_{\mathrm{i}}$ in the Lagrangian of the previous subsection. This corresponds to the $SU(2)\times U(1)$ version of the complex Ginzburg-Landau model~\cite{RevModPhys.65.851}.
The complex Ginzburg-Landau model is one of the simplest models that exhibits a synchronization phenomenon, and have a driven dissipative condensate (see Ref.~\cite{0034-4885-79-9-096001} for a review).
The MSR action has the form,
\begin{align}
\ri S&= \int {d^4x} \Bigl(
\ri \varphi_A^\dag(( -\partial_{0}^{2}+\bm{\nabla}^2-(2\ri\mu+\gamma)\partial_0-m^{2}_{\mathrm{r}}-\ri m^{2}_{\mathrm{i}})\varphi_R -2(\lambda_{\mathrm{r}}+\ri\lambda_{\mathrm{i}})(\varphi_R^\dag\varphi_R)\varphi_R)\notag\\
&\quad+
\ri \varphi_R^\dag((-\partial_{0}^{2}+\bm{\nabla}^2-(2\ri\mu-\gamma)\partial_0-m^{2}_{\mathrm{r}}+\ri m^{2}_{\mathrm{i}})\varphi_A -2(\lambda_{\mathrm{r}}-\ri\lambda_{\mathrm{i}})(\varphi_R^\dag\varphi_R))\varphi_A
-A\varphi_A^\dag\varphi_A
\Bigr).
\end{align}
Here,  we absorbed $\mu^{2}$ into the mass term. If we choose $m_{\mathrm{r}}^{2}=-\mu^{2}$, $m_{\mathrm{i}}=0$, $\lambda_{\mathrm{i}}=0$, 
this model reduces to the previous one. The term with the imaginary coupling $\ri\lambda_{\mathrm{i}}$ represents a nonlinear loss process.
Let us first see the fluctuation around the symmetric state with $\varphi_{R}=0$. We find the dispersion relation:
\begin{equation}
\begin{split}
\omega(\bm{k}) = \frac{1}{2}\Bigl(-\ri\gamma+2\mu\pm\sqrt{4 \ri( m_{\mathrm{i}}^2-  \gamma \mu) + 4 \mu^2 + 4 m_{\mathrm{r}}^2 - \gamma^2+\bm{k}^{2} }\Bigr).
\end{split}\label{eq:omegaforsymmetric}
\end{equation}
If $-m_{\mathrm{r}}^{2}$ or $m_{\mathrm{i}}^2$ is enough large,  the imaginary part of Eq.~(\ref{eq:omegaforsymmetric}) becomes positive and thus the symmetric state is unstable.
In this case, we need to find another stable state, which is obtained by the solution of the stationary condition $\delta S/\delta \varphi_R=0$.
We assume that the solution has the form $\phi_{R}= (0, v e^{-\ri\omega_{0}})$. Here, we introduced the synchronization frequency $\omega_{0}$,
which is necessary for the synchronization phenomenon as is seen in the following.
The stationary condition leads to 
\begin{equation}
\begin{split}
(\omega_{0}^{2}-2\mu\omega_{0}-m_{\mathrm{r}}^{2}-2\lambda_{\mathrm{r}}v^2+\ri(\gamma\omega_{0}- m_{\mathrm{i}}^{2} -2\lambda_{\mathrm{i}}v^2))v=0.
\label{eq:gapEq}
\end{split}
\end{equation}
If $m_{\mathrm{i}}^{2}=0$, and $\lambda_{\mathrm{i}}=0$, the nontrivial solution exists if $m_{\mathrm{r}}^{2}<0$ and $\lambda_{\mathrm{r}}>0$. In this case, we find $v= \sqrt{-m_{\mathrm{r}}^{2}/\lambda_{\mathrm{r}}}$ and $\omega_{0}=0$, i.e., there is no synchronization. This situation is nothing but that in the  previous example. In contrast, for the existence of $m_{\mathrm{i}}^{2}$ and $\lambda_{\mathrm{i}}$, the synchronization frequency is essential to obtain the solution is Eq.~\eqref{eq:gapEq}.
A remarkable point is that the nonvanishing condensation can occur even if $m_{\mathrm{r}}^{2}>0$, which is different from the condensation mechanism at the equilibrium system; the condensation is caused by the dissipation~\cite{0034-4885-79-9-096001}.
The explicit form of the solution is not important in our argument; we only assume its existence.

The Noether charges are the same as the previous ones, \eqref{eq:QAB} and \eqref{eq:QRB}. The expectation values of the commutators are obtained as
\begin{align}
 -\frac{1}{V}\langle[\ri Q_A^1,Q_R^2]\rangle=\frac{1}{V}\langle[\ri Q_A^2,Q_R^1]\rangle=(\mu-\omega_{0}) v^2. \label{eq:typeBCondition2}
\end{align}
That is, $Q_{A}^{1}$ and $Q_{A}^{2}$ belong to type-B NG fields. Next, let us consider fluctuations around the condensate.
By parametrizing the fields as $\varphi_R=((\chi^1_R+\ri \chi^2_R)e^{-\ri \omega_0 t}, (v+\psi^1_R+\ri \psi^2_R)e^{-\ri \omega_0 t})$ and $\varphi_A=((\chi^1_A+\ri \chi^2_A)e^{-\ri \omega_0 t},(\psi^1_A+\ri \psi^2_A)e^{-i \omega_0 t})$,
we find the  inverse propagators for the $\psi$ and $\chi$ sectors as
\begin{align}
D_{\psi R}^{-1}(\omega,\bm{k})&=
\begin{pmatrix}
-\omega^2-\ri \gamma\omega+\bm{k}^2+4\lambda_{r} v^{2} &  2\ri (\mu-\omega_{0})\omega\\\
-2\ri (\mu-\omega_{0})\omega +4\lambda_{\mathrm{i}}v^{2}  & -\omega^2-\ri \gamma\omega+\bm{k}^2
\end{pmatrix}, \label{eq:DinvpsiR}\\
D_{\chi R}^{-1}(\omega,\bm{k})&=
\begin{pmatrix}
-\omega^2-\ri \gamma\omega+\bm{k}^2 & 2\ri (\mu-\omega_{0})\omega\\
-2\ri (\mu-\omega_{0})\omega & -\omega^2-\ri \gamma\omega+\bm{k}^2
\end{pmatrix}.
\end{align}
Here, we note that the frequency $\omega$ is measured around that of the condensate $\omega_0$. 
At small $\bm{k}$, we obtain the diffusive NG mode (type A),
\begin{equation}
\begin{split}
\omega =-\ri \frac{|\bm{k}|^2}{\gamma+2(\mu-\omega_{0})\lambda_{\mathrm{i}}/\lambda_{\mathrm{r}}}
\end{split}
\end{equation}  in the $\psi$ sector, and the propagating mode (type-B), 
\begin{align}
\omega =  \frac{|\bm{k}|^2}{4(\mu-\omega_{0})^2+\gamma^2}(\pm2(\mu-\omega_{0})- \ri \gamma),
\end{align}
in the $\chi$ sector.
We emphasize that the behaviors of the dispersion relation for type-A and B NG modes do not change even in the nonequilibrium stationary state.

\section{Ward-Takahashi identity in dissipative systems}
\label{sec:ward}
In the previous section, we discussed the NG modes and their dispersion relations associated with the spontaneous breaking of $O(N)$ and $SU(2)\times U(1)$ models  in the saddle point approximation.
In this section, we nonperturbatively establish the result using a Ward-Takahashi identity.
We consider a system described by a Fokker-Planck equation, $\partial_t |P\rangle = -\hH_\text{FP} |P\rangle$.
We assume that the Fokker-Planck Hamiltonian does not explicitly depend on time, $\partial_t\hH_\text{FP}=0$, and the real part of right eigenvalues of $\hH_\text{FP}$ are non-negative and 
it contains at least one zero eigenvalue.
In general, a stationary state with the zero eigenvalue may not be a thermal state, i.e., a nonequilibrium steady state is allowed in this formalism.
We also assume that the stationary state does not break the spacetime symmetries.
We consider a continuum symmetry group $\mathcal{G}$ with a generator $\hQ^\alpha_A$ as an internal symmetry, which commutes with the Fokker-Planck Hamiltonian,  $[\hH_\text{FP}, \hQ^\alpha_A]=0$. For fields belonging to a linear representation, $\hphi_R^a$ and $\hphi_A^a$ transform as
\begin{align}
-[\ri \hQ_A^\alpha,\hphi_R^a] = \ri \T \hphi_R^b, \quad\text{and}\quad -[\ri \hQ_A^\alpha,\hphi_A^a] = \ri \T \hphi_A^b.
\end{align}
We also define $\hQ_R^\alpha$ such that 
\begin{align}
-[\ri \hQ_R^\alpha,\hphi_A^a] = \ri \T \hphi_R^b, \quad\text{and}\quad -[\ri \hQ_R^\alpha,\hphi_A^a] = 0.
\label{eq:commutationQR}
\end{align}
In examples of Sec.~\ref{sec:ssb}, the poisson bracket is defined. In these cases, the commutation relation of $\hQ_{R}$ coincides with the poisson bracket:
$  -[\ri \hQ_R^\alpha,\hphi_A^a]= \{ \hQ_R^\alpha,\hphi_R^a\}_{\text{PB}}$.
In general, $\hQ^\alpha_R$ does not commute with the Hamiltonian whereas it does in the Hamiltonian system.

In the following, we show the following relation from a Ward-Takahashi identity,
\begin{align}
[D^{-1}]^{ab}_{i j} (\omega=0, \bm{k}=0) \langle [\ri \hQ_A^\alpha, {\hvarphi}_j^a] \rangle=0, \label{eq:wardid}
\end{align}  
where $D_{i j}^{-1}$ is the inverse propagator and indices $i$ and  $j$ run $R$ and $A$. 
The derivation in this section is commonly used  in quantum field theories~\cite{WeinbergText}.

To drive Eq.~\eqref{eq:wardid}, it is useful to move to the path integral representation of the generating functional:
\begin{align}
Z[\bm{J}]=\int \D \bm{\phi} e^{i S[\bm{\phi}]+\ri \int {d^4x}  \bm{J}\cdot \bm{\vphi}},
\end{align} 
where we used the vector notation: $\bm{\phi}= (\phir^a,\phia^a)$ and  $\bm{J}=(J_A^a, J_R^a)$.
We assume that  $\vphi_i^a$ contains an order parameter, and
the action $S[\bm{\phi}]$ and the path integral measure $\D \bm{\phi}$ 
is invariant under the infinitesimal transformation,
$\bm{\phi} \rightarrow \bm{\phi}+ \epsilon_\alpha \delta_A^\alpha \bm{\phi}$.
Here, $\epsilon_\alpha$ is an infinitesimal parameter, and $\delta_A^\alpha {\bm{\hphi}}$= $-[\ri \hQ_A^\alpha, {\bm{\hphi}}]$ in the operator formalism. 
Since the generating functional is invariant under the reparameterization of the fields $\bm{\phi}(t,\bm{x})\to \bm{\phi}'(t,\bm{x})=\bm{\phi}(t,\bm{x})+\epsilon_\alpha \delta_A^\alpha \bm{\phi}(t,\bm{x})$, 
we have
\begin{align}
Z[\bm{J}]&=\int \D \bm{\phi}'e^{\ri S[\bm{\phi}']+\ri \int {d^4x}  \bm{J}\cdot \bm{\phi}'}\notag\\
&=\int \D \bm{\phi}e^{\ri S[\bm{\phi}]+\ri \int {d^4x}  \bm{J}\cdot \bm{\vphi}}
\biggl(1+\ri \epsilon_\alpha\int {d^4x}  \bm{J}(t,\bm{x}) \cdot \delta_A^\alpha \bm{\vphi}(t,\bm{x})\biggr) +\mathcal{O}(\epsilon^2) \notag\\
&= Z[\bm{J}] \biggl(1+ \ri \epsilon_\alpha \int {d^4x}  \bm{J}(t,\bm{x}) \cdot 
\langle \delta_A^\alpha \bm{\vphi}(t,\bm{x}) \rangle_J \biggr) +\mathcal{O}(\epsilon^2), \label{eq:generating}
\end{align}
where $\langle ... \rangle_J$ is the expectation value in the presence of the source $\bm{J}$: 
\begin{align}
\langle \bm{\vphi} \rangle_J  \equiv \frac{1}{Z[\bm{J}]}\int \D \bm{\phi}e^{i S[\bm{\phi}]+\ri \int {d^4x}  \bm{J}\cdot \bm{\vphi}} \bm{\vphi}.
\end{align}
From Eq.~\eqref{eq:generating}, we obtain the identity
\begin{align}
 \int {d^4x}  \bm{J} \cdot \langle [\ri \hQ^\alpha_A, {\bm{\hvarphi}}] \rangle_J=0,
\end{align}
where we used $\langle \delta_A^\alpha \bm{\vphi} \rangle_J=- \langle [\ri \hQ^\alpha_A, {\bm{\hvarphi}}] \rangle_J$.
Introducing the effective action 
\begin{align}
\Gamma [\bm{\vphi}] \equiv -\ri \ln Z[\bm{J}] - \int {d^4x}  \bm{\vphi}\cdot \bm{J},
\end{align}
we can write the identity as 
\begin{align}
\int {d^4x} \frac{\delta \Gamma}{\delta \vphi_j^b(t, \bm{x})} \langle [\ri \hQ^\alpha_A, {\hvarphi}_j^b(t, \bm{x})] \rangle_J=0, \label{eq:equality}
\end{align}
where we used $\delta\Gamma/\delta\bm{\phi} = -\bm{J}$.
Differentiating Eq.~\eqref{eq:equality} with respect to $\vphi^a_i(t',\bm{x}')$ and taking the limit of $J\to0$,
we arrive at
\begin{align}
\int {d^4x}  [D^{-1}]_{ab}^{i j} (t'-t, \bm{x}'-\bm{x})  \langle [\ri \hQ^\alpha_A, {\hvarphi}_j^b(t, \bm{x})] \rangle&=0,\label{eq:abababa}
\end{align}
where the inverse of the propagator is obtained as $ [D^{-1}]_{ab}^{i j}(t'-t, \bm{x}'-\bm{x})=\delta^2 \Gamma /\delta \vphi^a_i(t',\bm{x}') \delta \vphi^b_j(t,\bm{x})  $.
In the momentum space, the equation \eqref{eq:abababa} turns out to be Eq.~\eqref{eq:wardid}.
This identity represents the eigenvalue equation with the zero eigenvalue, whose eigenvectors  are $\langle [\ri \hQ_A^\alpha, {\hvarphi}_j^b] \rangle$.
The number of independent eigenvectors is equal to the number of broken Noether charges.
When $\hQ_R^\alpha$ is conserved, we obtain a similar result for $\hQ_R^\alpha$:
\begin{align}
[D^{-1}]_{ab}^{i j} (\omega=0, \bm{k}=0) \langle [\ri \hQ_R^\alpha, {\hvarphi}_j^b] \rangle=0. \label{eq:wardid3}
\end{align}  
For the $O(N)$ model with the expectation value $\langle \phi_R^a\rangle= \delta^{a}_1\phi_0$, we can write $-\langle[\ri \hQ^\alpha_A,\vphi_R^a]\rangle=\delta^{\alpha a}$ for $a=2,3,\dots, N$ by a useful normalization of $\hQ_A^\alpha$.
Then, from Eq.~\eqref{eq:wardid3}, we obtain 
\begin{align}
[D^{-1}]_{a \alpha}^{i R} (\omega=0, \bm{k}=0)=0. \label{eq:inversepropagator}
\end{align}
This identity gives a constraint to the dispersion relations. 

Now, we expand the inverse of the propagators with respect to $\omega$ and $\bm{k}$ as 
\begin{align}
[D^{-1}]_{ab}^{RR}&=0, \\
[D^{-1}]_{ab}^{AR} &=  \delta_{ab}C_{(0,2)}\bm{k}^2-\ri\delta_{ab}C_{(1,0)}\omega -\delta_{ab}C_{(2,0)}\omega^2
+\cdots,\\
[D^{-1}]_{ab}^{AA} &=-\ri \delta_{ab}A_{(0,0)}+\cdots,
\end{align} 
where the coefficients $C_{(n,m)}$ and $A_{(n,m)}$ are generally nonzero without any constraints from another symmetry.
Here, $[D^{-1}]_{ab}^{RR}$ vanishes due to conservation of the probability~\cite{tauber2007field}.
Furthermore, $C_{(0,0)}$ becomes zero from the constraint Eq.~\eqref{eq:inversepropagator}.
From the equation $[D^{-1}]_{ab}^{AR}(\omega,\bm{k})=0$, we obtained the dispersion relation for the dissipative NG modes $\omega\sim -\ri|\bm{k}|^2$.
In addition, when, $\Qr$ is also conserved, which corresponds to the Hamiltonian system, we find that $A_{(0,0)}$ vanishes from Eq.~\eqref{eq:wardid3}.
If the system satisfies the fluctuation-dissipation relation, $A_{(0,0)}$ is related to $C_{(1,0)}$: $A_{(0,0)}=2C_{(1,0)} T$ with the temperature $T$,
and thus, $C_{(1,0)}$ vanishes. 
In this case, the NG modes are the propagating modes with the dispersion relation $\omega= a_R|\bm{k}|-\ri a_I|\bm{k}|^2$~\cite{PhysRevD.91.056006},
where $a_R$ and $a_I$ are constants depending on $C_{(n,m)}$.
Even in the absence of the fluctuation-dissipation relation, we expect that the conservation of $\Qr$ leads to the vanishing $C_{(1,0)}$ since it represents a dissipation.  A more concrete proof will be given in our future work~\cite{minami2018}.

The symmetry breaking pattern discussed in the $O(N)$ model for $N\geq4$ is relatively a simple case because the broken Noether charges transform as the vector representation under  unbroken $O(N-1)$ symmetry. In other words, the NG modes belong to the vector representation of the $O(N-1)$.
The unbroken symmetry restricts couplings between NG modes and others, and then, the inverse of the propagators proportional to the Kronecker delta\footnote{When the unbroken symmetry is the antisymmetric tensor $O(2)$, there is a possibility to have $\epsilon^{ab}$ in the inverse of the propagator.}.
If the diffusive NG field has no internal unbroken charge, the analysis will be more complicated. In particular, the coupling between NG modes to hydrodynamic modes must be taken into account.

A different type of dispersion relations will be found when $\langle [\hQ_A^a,\hQ_R^b]\rangle$ is nonzero, which corresponds to Type-B modes.
For $SU(2)\times U(1)$ model, the unbroken symmetry is $U(1)$;
we have the two second rank invariant tensors in the real representation: the Kronecker delta $\delta^{ab}$ and the antisymmetric tensor $\epsilon^{ab}$.
Then, the inverse propagator is expanded as
\begin{align}
\begin{split}
[D^{-1}]_{ab}^{AR} &=  \delta_{ab}C^S_{(0,2)}\bm{k}^2+\epsilon_{ab}C^A_{(0,2)}\bm{k}^2-\ri\delta_{ab}C^S_{(1,0)}\omega-\ri \epsilon_{ab}C^A_{(1,0)}\omega +\cdots.
\end{split}
\end{align}
In this case, the dispersion relation has the form, $\omega = c_R|\bm{k}|^2-\ri c_I|\bm{k}|^2$, where $c_R$ and $c_I$ are constants. Therefore, the type-B NG mode can propagate.
We note that the dispersion relation of type-B NG modes in the Hamiltonian system is $\omega = b_R|\bm{k}|^2-\ri b_I|\bm{k}|^4$, where $b_R$ and $b_I$ are constants~\cite{PhysRevD.91.056006}.

For more general cases, the coefficients are matrices,
\begin{align}
\begin{split}
[D^{-1}]_{ab}^{AR} (\omega,\bm{k})&=  C_{(0,2)ab}\bm{k}^2-\ri C_{(1,0)ab}\omega-C_{(2,0)ab}\omega^2 +\cdots.
\end{split}
\end{align}
If $\det C^{ab}_{(1,0)}$ is nonzero, $C^{ab}_{(2,0)}$ is negligible at small $\bm{k}$, and we obtain the dispersion relation from the eigenvalue of $\ri [C^{-1}]^{ac}_{(1,0)}C_{(0,2)cb}\bm{k}^2$. The eigenvalue is generally complex, and thus, we have the form $\omega = d_R|\bm{k}|^2-\ri d_I|\bm{k}|^2$.
We expect the existence of the coefficient $d_R$ is related to the nonvanishing $\langle [\ri\hQ_A^a,\hQ_R^b]\rangle$, although we have not given a proof
which is beyond the scope of this paper.

\section{summary and discussion}
\label{sec:summary}
We discussed spontaneous symmetry breaking  and the Nambu-Goldstone (NG) modes in dissipative systems described by Langevin or  Fokker-Planck equations.
For this purpose, we employed the $O(N)$ and $SU(2)\times U(1)$ scalar models as toy models.
In the nondissipative limit, which corresponds to a Hamilton system, 
there exist the Noether charges $\Qr$ that are the generators of the internal symmetry by means of the Poisson bracket and they are conserved.
In contrast, $\Qr$ are no longer conserved due to dissipation and noise in the dissipative system.
Instead, there exist other conserved quantities $\Qa$, which are the Noether charges of the internal symmetry  in the Fokker-Planck equation.

The symmetry breaking is characterized by existence of a nonvanishing order parameter.
In the $O(N)$ model, $O(N)$ symmetry is spontaneously broken into $O(N-1)$, and the order parameter is $\langle [\ri \Qa, \hphir^a] \rangle$. Since $\langle[\ri Q_A^\alpha,Q_R^\beta]\rangle=0$, the NG modes belong to the type-A modes.  We found that the NG modes are the diffusive modes, $\omega\sim -\ri\bm{k}^2$.
This is the different behavior compared to the Hamiltonian system, where the NG modes are the propagating modes.
This difference is caused by whether $\Qr$ is conserved or not: When both $\Qa$ and $\Qr$ are conserved and their symmetry is spontaneously broken, there appear the propagating NG  modes. In the dissipative systems, when $\Qa$ is broken, the diffusive NG modes appear. We established this result by using the Ward-Takahashi identity for $\Qa$ and $\Qr$ symmetries in Sec.~\ref{sec:ward}.

We also discussed type-B NG modes in $SU(2)\times U(1)$ model, where $\langle [\ri \Qa, Q_R^\beta] \rangle \neq 0$. In this case, the dispersion relation of NG modes have the form of $\omega =a|\bm{k}|^2-\ri b|\bm{k}|^2$, while they are $\omega=a'|\bm{k}|^2-\ri b'|\bm{k}|^4$ in the Hamiltonian system, where  $a$, $b$, $a'$, and $b'$ are constant parameters.
In contrast to type-A NG modes, type-B NG modes are still propagating modes.
These different behaviors can be understood as the difference between the harmonic oscillation and precession motion.
In  Hamiltonian systems, there is one to one correspondence between the type A (B) and harmonic oscillation (precession motion) of NG modes~\cite{PhysRevLett.108.251602, PhysRevLett.110.091601, PhysRevX.4.031057,PhysRevD.91.056006}.
If one adds a small friction term into the equation of the harmonic oscillator, one will find a damped oscillation. If the friction is large, the motion turns to the overdamping motion. This is the case for the type-A mode. In contrast, if one adds a friction term into the equation for the precession motion, one will find a damped precession motion. However, this motion is never overdamped. This phenomenon is also observed in this paper for the type-B mode.

In this paper, we focus only on classical systems. Generalization to quantum systems is straight forward: We may add higher terms in $\phi_A$ such as $(\phi^b_A)^2\phi^a_A\phi^a_R$, and take into account the Bose and Fermi statistics.
This symmetry breaking pattern of $SU(2) \times U(1)\to U(1)$ is a similar to the spinor BEC, where the symmetry breaking pattern is $SO(3)\times U(1)\to U(1)$~\cite{RevModPhys.85.1191, KAWAGUCHI2012253}.
It is interesting that this driven dissipative condensate and type-B NG modes discussed in this paper is observed in open quantum systems.

Our approach can apply to the spontaneous breaking of spacetime symmetries, although  our result in this paper is limited to that of internal symmetry.
Even in the Hamiltonian system, the general counting rule and  dispersion relation of their NG modes have not been well-understood.
An interesting example in a dissipative system is discussed in the active hydrodynamics, where energy and momentum are not conserved, but equations of motion respect spacetime translational and rotational symmetries~\cite{PhysRevLett.75.4326, PhysRevE.58.4828}.
In this situation, the velocity fields is the order parameter and it breaks the rotational symmetry. For $d$-spatial dimensions, there appear $d-2$ diffusive (shear) modes,
and one propagating sound mode. This sound mode caused by the mixing between longitudinal NG mode and the hydrodynamic mode associated with the 
number conservation~\cite{PhysRevLett.75.4326, PhysRevE.58.4828}. The mixing of hydrodynamic mode can change the dispersion relation.

 It is interesting to clarify the relation between the broken symmetry, the NG modes and their dispersion relations in dissipative systems.
Recently, in Hamiltonian systems without the Lorentz invariance,
those relations have been made clear~\cite{PhysRevLett.108.251602, PhysRevLett.110.091601, PhysRevX.4.031057, PhysRevD.91.056006},
which does not cover the dissipative system.
In this paper, we generalized the theorem in the Hamiltonian systems to that in dissipative systems.
From our observations, we propose the following conjecture:
We suppose that a Fokker-Planck Hamiltonian $H$ commutes with a generator $Q_A^\alpha$ of a Lie group $\mathcal{G}$.
We also suppose that a Poisson bracket is defined, and $Q_R^\alpha$ exists as a generator of $\mathcal{G}$ in the sense of the Poisson bracket.
In general, $Q_R^\alpha$ does not commute with the Fokker-Planck Hamiltonian.
When the $\mathcal{G}$ is spontaneously broken into its subgroup $\mathcal{H}$, the number of type-A ($N_\text{A}$) and type-B ($N_\text{B}$) NG modes will be given as 
\begin{align}
N_\text{A}&=N_\text{BS}-\mathrm{rank}\langle [\ri Q_A^\alpha, Q_R^\beta]\rangle,\\
N_\text{B}&=\frac{1}{2}\mathrm{rank}\langle [\ri Q_A^\alpha, Q_R^\beta]\rangle.
\end{align}
where $N_\text{BS}=\mathrm{dim}(\mathcal{G}/\mathcal{H})$ is the number of broken symmetries.
These equations correspond to Eq.~\eqref{eq:NGrelation} for the Hamiltonian system. 
Their dispersion relations will be classified into four types:
\begin{align}
&\text{Type-A}\left\{
\begin{array}{ll}
\omega &= c_R|\bm{k}|-\ri c_I|\bm{k}|^2,\,\, \text{ if $[Q_R^\alpha,H]=0$},\\
\omega &= -\ri c_I|\bm{k}|^2, \,\,\text{\quad \qquad if $[Q_R^\alpha,H]\neq0$},
\end{array}
\right.\\
&\text{Type-B}\left\{
\begin{array}{ll}
\omega &= c_R|\bm{k}|^2-\ri c_I|\bm{k}|^4, \text{ if $[Q_R^\alpha,H]=0$},\\
\omega &= c_R|\bm{k}|^2-\ri c_I|\bm{k}|^2, \text{ if $[Q_R^\alpha,H]\neq0$}.
\end{array}
\right.
\end{align}
Of course, the models discussed in this paper satisfy these relations.
We leave the detailed analysis and a proof of this conjecture leave to our future work~\cite{minami2018}.
 
\acknowledgements
Y.M. thanks K.~Uriu, T.~Okada, K.~Sato, Y.~Tanizaki, N.~Ogawa, M.~Hongo, S.~Iso, and K.~Itakura for useful discussion.
This work was partially supported by Japan Society of Promotion of Science (JSPS), Grants-in-Aid for Scientific Research
(KAKENHI) Grants No. 15H03652, 16K17716, and 17H06462, and by RIKEN iTHES Project and iTHEMS Program.
\appendix

\section{Symmetry of Hamiltonian and dissipative systems in the real-time formalism}
\label{sec:keldysh}

Here, we discuss symmetry of an action in Hamiltonian and dissipative systems from the real-time formalism.
To construct the dissipative system, we couple the system with an environment, and integrate out the environment fields. 
We shall see that the doubling of the symmetry, $\Qmr$ and $\Qma$, occurs in the real-time formalism and the coupling to the environment explicitly violates the $\Qmr$ symmetry.

We consider the following Lagrangian to discuss the symmetry of the Hamiltonian system,
\begin{align}
\mathcal{L}=\frac{1}{2}(\partial_t\phi^a)^2-\frac{1}{2}(\bm{\nabla}\phi^a)^2-\frac{1}{2}m^2(\phi^a)^2
 - \frac{u_0^2}{4}((\phi^a)^2)^2.
 \label{eq:original}
\end{align}
which is invariant under $\phi^a \rightarrow \phi^a+ \ri \epsilon_\alpha \T \phi^b$.
\begin{figure}
\centering
    \includegraphics[width=0.5\hsize]{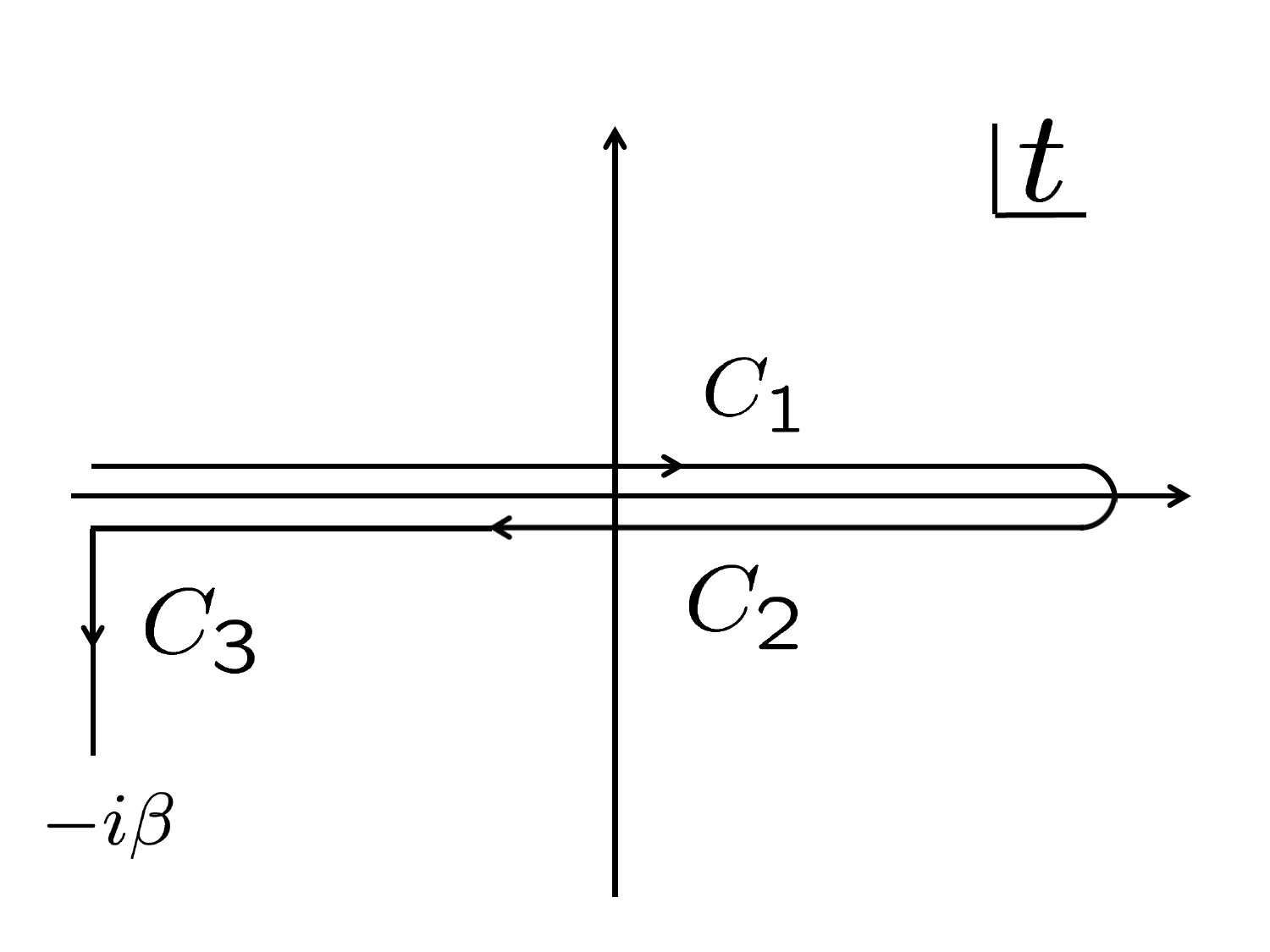}
  \caption{The contour on the complex time plane in the real-time formalism.
  The branch $C_1$ runs on the real axis from $-\infty$ to $\infty$, 
  while $C_2$ runs backward from $\infty$ to $-\infty$.
  $C_3$ runs from $-\infty$ to $-\infty-\ri\beta$, where $\beta$ is the inverse temperature.}
 \label{fig:contour}    
\end{figure}
The generating functional in the real-time formalism is expressed as the path integral on the complex time-path shown in Fig.~\ref{fig:contour},
which is 
\begin{align}
Z[J^1_a, J^2_a]=&\int \D \phi_1^a \D \phi_2^a e^{ \ri  S[\phi_i^a]
+ \ri \int {d^4x}  J^i_a \phi^a_i },
\end{align}
with the action,
\begin{align}
 S[\phi_1^a, \phi_2^a]=&-\frac{1}{2}\int {d^4x} {d^4x'} 
\phi_i^a (x) [D^{-1}]^{ij}(x-x')  \phi_j^a(x') \notag \\
&- \frac{u_0^2}{4}\int {d^4x} 
\biggl( ((\phi^a_1(x))^2)^2-((\phi^a_2(x))^2)^2\biggr),
\end{align}
where the subscript $i, j$ run $1, 2$, and $\phi_{1,2}^a$ represent the fields on the forward and backward branches, respectively
(See the caption of Fig.~\ref{fig:contour}).
Here, $[D^{-1}]^{i j}$ are the inverse of the propagators, which are given as~\cite{le2000thermal}
\begin{align}
-\ri D_{11} (\omega, \bm{k})
&=\ri\bm{\mathrm{P}}\frac{1}{\omega^2-{\bm{k}^2}{-}m^2} +\Bigl(\frac{1}{2}+n(\omega)\Bigr)\rho(\omega,\bm{k}),\\
-\ri D_{22}(\omega, \bm{k})&=(-\ri D_{11}(\omega, \bm{k}))^*, \\
-\ri D_{12}(\omega, \bm{k}) &=n(\omega)\rho(\omega,\bm{k}),\\
-\ri D_{21}(\omega, \bm{k}) &=(1+n(\omega))\rho(\omega,\bm{k}),
\end{align}
where $\bm{\mathrm{P}}$ denotes the principal value, $n(\omega)=1/(e^{\beta\omega}-1)$ the Bose distribution function, 
and $\rho(\omega,\bm{k})=2\pi \varepsilon(\omega)\delta(\omega^2-\bm{k}^2-m^2)$ the spectral function
with the sign function $\varepsilon(\omega)$.
The point in the real time formalism is that the doubling of the fields occurs: 
$\phi^a \rightarrow \phi_1^a$ and $\phi^a_2$. 
This doubling causes the doubling of symmetry.

Before discussing the symmetry, we change the field variables to 
\begin{align}
\phir^a &= \frac{1}{2}(\phi_1^a+ \phi_2^a), \quad \phia^a =\phi_1^a-\phi_2^a. \label{eq:keldyshrotation}
\end{align}
The generating functional and the action for $\phi_{R,A}^a$ is written in the new variables as
\begin{align}
Z[J^R_a, J^A_a]=&\int \D \phi_R^a \D \phi_A^a e^{ \ri S[\phir,\phia]
+ \ri \int {d^4x}  (J_a^{R} \phir^a +J_a^{A} \phia^a) },\\
  S[\phir^a, \phia^a]=&-\frac{1}{2}\int{d^4x} {d^4x'} 
\phi_i^a (x)  [D^{-1}]^{ij}(x-x')  \phi_j^a(x') \notag \\
&-  u_0^2\int {d^4x} 
\biggl( (\phir^a)^2 \phir^b \phia^b +\frac{1}{4}(\phia^a)^2 \phir^b{\phia^{b}}  \biggr), \label{eq:closedaction}
\end{align}
where the subscript $i$ and  $j$ run $R$ and $A$, and $J^R_a \equiv J^1_a +J^2_a$ and
$J^A_a \equiv (J^1_a -J^2_a)/2$. 
If we drop $(\phia^a)^2 \phir^b{\phia^{b}}/4$ term, the potential term reduces to that of classical field theory~\eqref{eq:action}.  
{
Here, $D_{R, A}^{-1}$ is given in momentum space as
\begin{align}
[D^{-1}]^{ij}(\omega,\bm{k})=  \begin{pmatrix}
0 &  D_A^{-1}(\omega,\bm{k})\\
 D_R^{-1}(\omega,\bm{k}) & \Bigl(\frac{1}{2}+n(\omega)\Bigr)(D_R^{-1}(\omega,\bm{k})-D_A^{-1}(\omega,\bm{k}))
\end{pmatrix},
\end{align}
where $D_R^{-1}(\omega,\bm{k})\equiv -(\omega+\ri \epsilon)^2+\bm{k}^2+m^2$ and $D_A^{-1}(\omega,\bm{k})=(D_R^{-1}(\omega,\bm{k}))^*$.
}
 $[D^{-1}]^{AA} $ is infinitesimally small because
\begin{align}
[D^{-1}]^{AA} &=  \Bigl(\frac{1}{2}+n(\omega)\Bigr)\bigl( D_{R}^{-1}-D_{A}^{-1}\bigr) \notag \\
&= -4\ri \epsilon \omega\Bigl(\frac{1}{2}+n(\omega)\Bigr)  \sim \epsilon.
\end{align}

Let us now discuss the symmetry of Eq.~\eqref{eq:closedaction}.
We can easily see that the action is invariant under the following transformation:
\begin{align}
\phia^a \rightarrow \phia^a + \ri {\epsilon_{\alpha} }\T \phia^b, \quad 
\phir^a \rightarrow \phir^a + \ri {\epsilon_{\alpha}} \T \phir^b. \label{eq:Qasymmtry}
\end{align} 
The Noether charge of this symmetry is written as
\begin{align}
\Qa = - \int d^3{x} \biggl[ \hpia^a \ri \T \hphir^b + \hpir^a \ri \T \hphia^b \biggr], \label{eq:Qa}
\end{align}
where the $\pi_{R, A}^a$  are defined as
\begin{align}
\pir^a \equiv (\pi_1^a+\pi_2^a)/2, \quad \pia^a \equiv \pi_1^a -\pi_2^a.
\end{align}
Here, $\pi_{1,2}^a=\partial_t{\phi}_{1,2}^a$ are the canonical momentum of $\phi_{1, 2}^a$ and satisfy the commutation relations
\begin{align}
[\hphi_1^a(t,\bm{x}), \hpi_1^b(t,\bm{y})]=\ri  \delta^{ab}\delta^{(3)}(\bm{x}-\bm{y}), \quad 
[\hphi_2^a(t,\bm{x}), \hpi_2^b(t,\bm{y})]=-\ri  \delta^{ab}\delta^{(3)}(\bm{x}-\bm{y}).
\end{align}
We note that $\phi_2^a$ and $\pi_2^a$ are the fields on the backward branch $C_2$
hence the commutation relation has the negative sign.
We can see that the form of the charge~\eqref{eq:Qa} is equal to that of the Langevin equation \eqref{eq:Qma}.

Furthermore, in the limit $\epsilon\to0$, Eq.~\eqref{eq:closedaction} is invariant under the transformation,
\begin{align}
\phia^a \rightarrow \phia^a + \ri \epsilon_{\alpha} \T \phir^b, \quad 
\phir^a \rightarrow \phir^a + \frac{\ri}{4}\epsilon_{\alpha} \T \phia^b, \label{eq:Qrsymmetry}
\end{align} 
and its Noether charge is given as
\begin{align}
\Qr =  -\int d^3{x} \biggl[ \hpir^a \ri \T \hphir^b + \frac{1}{4}\hpia^a \ri \T \hphia^b \biggr].
\label{eq:Qr}
\end{align}
This infinitesimally small breaking term plays an important role in the dissipation. This $\epsilon$ must be taken the zero limit after the infinite volume limit because these limits are not commutative. In other words, $\Qr$ is ``spontaneously'' broken. This charge corresponds to the charge~\eqref{eq:Qmr} if we drop $(1/4)\pia^a \ri\T \phia^b$ term in Eq.~\eqref{eq:Qr}. In this case, this is the symmetry of the action~\eqref{eq:closedaction} if $(\phia^a)^2 \phir^b{\phia^{b}}/4$ term is dropped.

We have seen that the action of the Hamiltonian system~\eqref{eq:closedaction} is invariant under the transformations by $\Qr$ and $\Qa$.
In the original Lagrangian~\eqref{eq:original} is invariant under the transformation $\phi^a \rightarrow \phi^a+ \ri \epsilon_\alpha \T \phi^b$.
Meanwhile, the action of the real-time formalism~\eqref{eq:closedaction} is invariant under the two transformations, Eqs.~\eqref{eq:Qa} and \eqref{eq:Qr}.

Next, we consider the system of $\phi_i^a$ coupled with other environment scalar fields $\Phi_i^{a}$.
We assume that the interaction between $\phi_i^a$ and $\Phi_i^{a}$ has the following form:
\begin{align}
 S_\text{int}[\phi_i^a,\Phi_i^{a}] = g\int d^4x(\phi_A^a\Phi^{a}_R +\phi_R^a\Phi^{a}_A),
\end{align}
where $g$ is the coupling constant.
Then the total action is ${S}_{\text{total}} = {S}[\phi_i^a] +  {S}[\Phi_i^{a}]+{S}_\text{int}[\phi_i^a,\Phi_i^{a}]$, where
\begin{align}
 S[\Phi_i^{a}]& =  -\frac{1}{2}\int {d^4x} {d^4x'}
\begin{pmatrix}
\Phi_R^{a} & \Phi_A^{a}
\end{pmatrix}
\begin{pmatrix}
0 &  G_{A}^{(0)-1} \\
 G_{R}^{(0)-1} &  [G^{(0)-1}]^{AA}  
\end{pmatrix}
\begin{pmatrix}
\Phi_R^{a} \\ \Phi_A^{a} 
\end{pmatrix}
-  V[\Phi_i^a],
\end{align}
with   potential term $V[\Phi_i^a]$, and
\begin{align}
G_{R, A}^{(0)-1}(\omega ,\bm{k}) &=-( \omega\pm \ri \epsilon)^2 +\bm{k}^2+M^2, \quad
[G^{(0)-1}]^{AA}(\omega, \bm{k}) =  \Bigl(\frac{1}{2}+n(\omega)\Bigr) \biggl(G_{R}^{(0)-1}-G_{A}^{(0)-1} \biggr)
\end{align}
in momentum space.

By integrating the environment fields $\Phi_i^{a}$ out, we obtain the effective action for $\phi_i^a$:
\begin{align}
 S_\text{eff} &=-\frac{1}{2}\int{d^4x} {d^4x'}
 \begin{pmatrix}
\phir^a & \phia^a
\end{pmatrix}
\begin{pmatrix}
0 &  D_A^{-1} - g^2 {G}_{A}\\
 D_R^{-1}- g^2 {G}_{R} &   [D^{-1}]^{AA}  - g^2 {G}_K  
\end{pmatrix}
\begin{pmatrix}
\phir^a \\ \phia^a
\end{pmatrix}\notag\\
&\qquad - u_0^2
\int {d^4x}
\biggl( (\phir^a)^2\phir^b\phia^b +\frac{1}{4}(\phia^a)^2\phir^b\phia^b \biggr) +O(g^3),
\label{eq:openaction}
\end{align}
where $G_{R,A}$ are the dressed Green function, and  ${G}_K$ is
\begin{align}
{G}_K(\omega, \bm{k}) &= \Bigl(\frac{1}{2}+n(\omega)\Bigr)\biggl({G}_{R}(\omega,\bm{k}) - {G}_{A}(\omega,\bm{k})\biggr).
\label{eq:GK}
\end{align}
We now discuss the symmetry of the action~\eqref{eq:openaction}.
The point is that it is not invariant under Eq.~\eqref{eq:Qrsymmetry} because
${G}_K(\omega,\bm{k}) \neq 0$; see Eqs.~\eqref{eq:openaction} and \eqref{eq:GK}.
In this sense, the $\Qr$ symmetry~\eqref{eq:Qrsymmetry} is broken. 
In contrast, $\Qa$ symmetry remains even in the open or dissipative system. 

\bibliography{ssb}

\end{document}